\documentclass[manuscript,screen,bookmarksnumbered,unicode,hidelinks]{acmart}

\AtBeginDocument{%
  }

\setcopyright{none}
\acmJournal{TOSEM}
\acmDOI{}
\acmVolume{0}
\acmNumber{0}
\acmArticle{0}
\acmMonth{0}

\usepackage{braket}
\usepackage{booktabs}
\usepackage{url}
\usepackage{xurl}
\usepackage{enumitem}
\usepackage{tcolorbox}
\tcbuselibrary{listings}
\usepackage{listings}
\usepackage{tabularx}
\usepackage{placeins}

\lstdefinestyle{artifactbash}{
  basicstyle=\ttfamily\footnotesize,
  columns=fullflexible,
  keepspaces=true,
  frame=none,
  showstringspaces=false,
  breaklines=true,
  breakatwhitespace=false
}

\newtcblisting{cmdbox}{
  listing only,
  colback=gray!8,
  colframe=black!35,
  arc=1mm,
  boxrule=0.4pt,
  left=1mm,
  right=1mm,
  top=1mm,
  bottom=1mm,
  listing options={style=artifactbash}
}

\begin{document}

\title{C2$\ket{\mathrm{Q}}$: A Robust Framework for Bridging Classical and Quantum Software Development --- RCR Report}

\author{Boshuai Ye}
\email{boshuai.ye@oulu.fi}
\orcid{0009-0000-3480-1234}
\affiliation{%
  \institution{M3S Research Group, SEIS Unit, University of Oulu}
  \city{Oulu}
  \country{Finland}
}

\author{Arif Ali Khan*}
\email{arif.khan@oulu.fi}
\orcid{0000-0002-8479-1481}
\affiliation{%
  \institution{M3S Research Group, SEIS Unit, University of Oulu}
  \city{Oulu}
  \country{Finland}
}

\author{Teemu Pihkakoski}
\email{teemu.pihkakoski@oulu.fi}
\orcid{0009-0008-4598-7271}
\affiliation{%
  \institution{Nano and Molecular Systems Research Unit, University of Oulu}
  \city{Oulu}
  \country{Finland}
}

\author{Peng Liang}
\email{liangp@whu.edu.cn}
\orcid{0000-0002-2056-5346}
\affiliation{%
  \institution{School of Computer Science, Wuhan University}
  \city{Wuhan}
  \country{China}
}

\author{Muhammad Azeem Akbar}
\email{azeem.akbar@lut.fi}
\orcid{0000-0002-4906-6495}
\affiliation{%
  \institution{Software Engineering Department, Lappeenranta-Lahti University of Technology}
  \city{Lappeenranta}
  \country{Finland}
}

\author{Matti Silveri}
\email{matti.silveri@oulu.fi}
\orcid{0000-0002-6319-2789}
\affiliation{%
  \institution{Nano and Molecular Systems Research Unit, University of Oulu}
  \city{Oulu}
  \country{Finland}
}

\author{Lauri Malmi}
\email{lauri.malmi@aalto.fi}
\orcid{0000-0003-1064-796X}
\affiliation{%
  \institution{Department of Computer Science, Aalto University}
  \city{Espoo}
  \country{Finland}
}

\renewcommand{\shortauthors}{Ye et al.}

\begin{abstract}
This is the Replicated Computational Results (RCR) Report for the paper \textit{C2$\ket{\mathrm{Q}}$: A Robust Framework for Bridging Classical and Quantum Software Development}. The paper introduces a modular, hardware-agnostic framework that translates classical problem specifications---Python code or structured JSON---into executable quantum programs across ten problem families and multiple hardware backends. We release the framework source code on GitHub at \url{https://github.com/C2-Q/C2Q}, a pretrained parser model on Zenodo at \url{https://zenodo.org/records/19061125}, evaluation data in a separate Zenodo record at \url{https://zenodo.org/records/17071667}, and a PyPI package at \url{https://pypi.org/project/c2q-framework/} for lightweight CLI and API use. Experiment~1 is supported through a released pretrained model and training notebook, while Experiments~2 and~3 are directly executable via documented \texttt{make} targets. This report describes the artifact structure, setup instructions, and the mapping from each execution route to the corresponding experiment.
\end{abstract}

\maketitle

\section{Overview}
\label{sec:overview}

This RCR report accompanies the TOSEM paper \textit{C2$\ket{\mathrm{Q}}$: A Robust Framework for Bridging Classical and Quantum Software Development}~\cite{ye2025c2}. It describes the released artifact, the execution routes supported by the artifact, and the extent to which the reported computational evidence can be independently regenerated.

\subsection{Paper Summary}
\label{sec:paper_summary}

Developing quantum software still requires expertise that most classical engineers do not have---from encoding problems into circuits to selecting hardware and interpreting raw measurement outputs~\cite{ye2025c2}. C2$\ket{\mathrm{Q}}$ is designed to bridge this gap. Given a classical problem description, either as a Python code snippet or as a structured JSON specification, the framework automatically transforms the classical problem description into a quantum result by classifying the problem, generating a Quantum-Compatible Format (QCF), constructing and transpiling circuits, recommending a suitable backend, executing the generated quantum program, and decoding the output into a user-readable format. The pipeline is organised into three modules: \textit{encoder}, \textit{deployment}, and \textit{decoder}, and currently covers ten problem families: combinatorial optimisation (MaxCut, MIS, TSP, Clique, K-Coloring, Vertex Cover), integer factorisation, and arithmetic operations (addition, subtraction, multiplication).

\textbf{Key claims.} The paper is structured around three research questions (RQs)~\cite{ye2025c2}:

\begin{itemize}[leftmargin=*,itemsep=3pt]
  \item \textbf{RQ1:} \textit{How can a framework be designed to bridge the gap between classical and quantum software development?}
  This research question is further decomposed into two sub-questions:
  \begin{itemize}[leftmargin=1.5em,itemsep=2pt]
    \item \textbf{RQ1.1:} \textit{What is an efficient approach for analysing classical code snippets and transforming them into quantum-compatible formats (QCFs) that can serve as input to quantum algorithms?}
    \item \textbf{RQ1.2:} \textit{How can suitable quantum algorithms be automatically selected and configured based on the translated QCFs to produce executable quantum programs?}
  \end{itemize}

  \item \textbf{RQ2:} \textit{How can the framework automatically recommend and deploy quantum hardware that best matches a program's requirements while hiding platform-specific complexity?}

  \item \textbf{RQ3:} \textit{How effective and generalisable is the proposed framework?}
\end{itemize}

\textbf{Key results.} The three RQs are addressed through three experiments. Quantitative findings are reported in the original paper~\cite{ye2025c2}; we summarise each experiment and point to the relevant sections below.

\begin{itemize}[leftmargin=*,itemsep=2pt]
  \item \textit{Experiment~1 -- Encoder evaluation (\textbf{RQ1}):} Evaluates the parser on a labeled dataset of synthetic Python programs and structured JSON inputs, reporting per-class precision, recall, and F1-score, as well as an end-to-end encoder completion rate covering data extraction, QCF translation, and circuit generation. The training process and recorded results are documented in the notebook \path{src/parser/parser_train_results_12_1.ipynb} in the GitHub repository~\cite{C2Q_github_2025}; the released pretrained model and training notebook together provide the main publicly accessible evidence for this experiment. Results are reported in Section~5.2 and Table~11 of~\cite{ye2025c2}.

  \item \textit{Experiment~2 -- Deployment evaluation (\textbf{RQ2}):} Evaluates the hardware recommender by running QAOA circuits for MaxCut on 3-regular graphs across nine quantum devices from five providers (IBM, IonQ, IQM, Rigetti, Quantinuum), reporting estimated error rates, execution times, and costs as a function of qubit count (scaling up to 56 qubits), and the device selected under default and varied scoring weights. This experiment does not require the parser model and is fully executable from the released artifact. Results are reported in Section~5.3 and Table~12 of~\cite{ye2025c2}.

    \item \textit{Experiment~3 -- Full workflow validation (\textbf{RQ3}):} Executes the complete pipeline on all benchmark inputs---434 Python programs and 100 JSON problem instances using Qiskit Aer, with a representative instance additionally validated on real hardware (IBM Brisbane and Finland's Helmi). Both a Python code input route (requiring the parser model) and a JSON input route (model-free) are supported. Reports the end-to-end workflow completion rate and a proxy-based usability comparison, handwritten lines of code, and explicit configuration decisions, against manual Qiskit implementations. The simulator-based workflow is fully executable from the released artifact, whereas the real-hardware portion serves as an access-dependent validation case and is therefore documented through archived outputs rather than positioned as an independently repeatable route. Results are reported in Section~5.4 and Table~14 of~\cite{ye2025c2}.
\end{itemize}

\subsection{Replication Process Overview}
\label{sec:replication_scope}

This report documents three practical execution routes: a Docker-based preliminary validation path (Section~\ref{sec:docker_path}), a source-checkout path for reproducing the paper experiments (Sections~\ref{sec:primary_path}, \ref{sec:model_setup}, and \ref{sec:reproduction_commands}), and a lightweight PyPI functional route (Section~\ref{sec:pypi_install}). Model-free JSON runs can be completed before installing the parser model, which is needed only for the Python code-input path (Section~\ref{sec:model_setup}). Full paper reproduction targets and their expected outputs are documented in Section~\ref{sec:reproduction_commands}.

\section{Artifact}
\label{sec:artifact}

The artifact is distributed through three public channels. The parser model is provided as a separate archive because it is a large trained checkpoint. Table~\ref{tab:artifacts} lists the main components with their locations, licenses, and roles, organised by their role in the artifact evaluation.

\begin{table}[!ht]
\centering
\caption{Main components of the released C2$\ket{\mathrm{Q}}$ artifact. All components are released under open licenses.}
\label{tab:artifacts}
\begin{tabularx}{\columnwidth}{p{2.2cm} p{2.0cm} p{1.5cm} X}
\toprule
\textbf{Component} & \textbf{Location} & \textbf{License} & \textbf{Purpose} \\
\midrule
Framework source & GitHub~\cite{C2Q_github_2025} & Apache-2.0 &
Maintained source, \texttt{make} targets, and documentation; primary evidence-generating mechanism \\

Evaluation data \& outputs & Zenodo data record~\cite{ye_khan_c2q_dataset_2025} & CC-BY-4.0 &
Benchmark inputs, archived reports, and recommender outputs; generated evidence \\

Validation scripts \& notebooks & GitHub; Zenodo data record~\cite{ye_khan_c2q_dataset_2025} & Apache-2.0 / CC-BY-4.0 &
Tests, analysis notebooks, and supporting scripts \\

Parser model & GitHub release; Zenodo model record~\cite{ye_parser_model_2025} & Apache-2.0 &
Pretrained CodeBERT model for Python code-input execution \\

PyPI package & PyPI~\cite{c2q_pypi_2025} & Apache-2.0 &
Lightweight CLI (\texttt{c2q-json}) and Python API; model-free out of the box, with parser-backed use requiring a separately downloaded model; supports Python~3.12--3.13; does not support \texttt{make}-based reproduction workflow \\

\bottomrule
\end{tabularx}
\end{table}
The archival components of the artifact are available on Zenodo with the following DOIs: evaluation data and archived outputs, 10.5281/zenodo.17071667; parser model, 10.5281/zenodo.19061126.


The GitHub repository is the primary executable artifact and the only channel that supports the \texttt{make}-based reproduction targets described in this report, including both the source-checkout and Docker routes. Zenodo preserves stable archival copies of the evaluation data and the parser model, while PyPI provides a lightweight command-line entry point for quick functional checks.

The packaged artifact also includes the required documentation files---\texttt{README.md}, \texttt{REQUIREMENTS.md}, \texttt{STATUS.md}, \texttt{LICENSE}, and \texttt{INSTALL.md}---covering acquisition, setup, environment assumptions, licensing, and a basic validation example. These files are available in the root of the GitHub repository~\cite{C2Q_github_2025}. The PyPI project page provides a lightweight installation entry point and links back to these GitHub-hosted companion documents; the GitHub repository remains the authoritative documentation location for artifact evaluation.

For the recommender evaluation, device metadata and coupling maps are included as archived artifact inputs. This freezes the snapshot used in the paper and makes the recommender outputs comparable across reproduction runs. New devices can be supported by adding corresponding metadata and coupling-map entries to the recommender module.

At a high level, the released materials can be grouped into three categories:

\begin{itemize}[leftmargin=*,itemsep=2pt]
  \item \textbf{Evidence-generating mechanisms}, including the framework implementation, benchmark inputs, JSON DSL specifications, Python programs, and the pretrained parser model used for Python code-input runs. Together, these constitute the main executable basis for regenerating the released results. For Experiment~1, the released model primarily supports inspection and inference-level validation, while the training workflow is documented separately through the notebook materials.

  \item \textbf{Generated evidence}, including archived benchmark reports, recommender outputs, and validation results preserved in the Zenodo data record~\cite{ye_khan_c2q_dataset_2025}. These materials enable direct comparison against locally regenerated outputs without requiring evaluators to rerun the full workload.

  \item \textbf{Data-analysis materials}, including the parser training notebook, dataset diversity analysis scripts, post-processing notebooks, visualisation materials, and the proxy-based usability analysis resources archived in the Zenodo data record~\cite{ye_khan_c2q_dataset_2025}.
\end{itemize}

\section{Prerequisites and Requirements}
\label{sec:requirements}

\subsection{Prerequisites}
\label{sec:skills}

Users of this artifact are expected to be comfortable with Python, standard command-line tools, virtual environments, \texttt{pip}, and basic \LaTeX{} for generating PDF reports. No background in quantum programming is needed to follow the documented routes.

\subsection{System Requirements}
\label{sec:system_requirements}

\textbf{Python and dependency versions.} The source-checkout and PyPI routes are validated on Python~3.12 and Python~3.13. The package metadata declares \texttt{>=3.12,<3.14}. Python~3.14 and newer are outside the validation scope of this release because the artifact depends on binary scientific, parser-model, and quantum SDK packages whose Python~3.14 support is not yet part of the reviewed dependency profile. A local interpreter can be checked with \texttt{python3.12 --version} or \texttt{python3.13 --version} on macOS/Linux, and with \texttt{py -3.12 --version} or \texttt{py -3.13 --version} on Windows. If a local Python~3.12/3.13 environment is unavailable, the Docker route provides a model-free validation path. The source-checkout route also requires standard Python package-management tools and a working \texttt{pdflatex} installation for report generation.  The artifact uses pinned dependencies to keep reproduction runs stable. The current release pins \texttt{qiskit==2.5.0} and \texttt{qiskit-aer==0.17.2}, together with a NumPy~2-compatible scientific stack. These pins define the validated reproduction environment.

\textbf{Setup time.} Under normal network and package-installation conditions, cloning the repository, creating the virtual environment, and installing dependencies typically takes under 30 minutes on a clean machine. Downloading and unpacking the parser model adds a few minutes on top for the Python code-input route. The smoke targets complete in minutes and are the recommended first check. Full-scale reproduction runs are intentionally not part of the installation check; their runtimes are noted in Section~\ref{sec:reproduction_commands}.

\subsection{Parser Model Dependency}
\label{sec:model_dependency}
Running the framework on Python code inputs requires a pretrained parser model; JSON inputs work without it. The model is distributed separately because it is a large trained checkpoint. Keeping the parser model outside the GitHub source tree and PyPI package keeps installation lightweight while preserving the model as a citable archival artifact. By default, the repository expects the model under \path{src/parser/saved_models_2025_12/}. The model archive is available as a GitHub release asset\footnote{\url{https://github.com/C2-Q/C2Q/releases/download/v1.0-artifact/saved_models_2025_12.zip}} and as a Zenodo archival copy~\cite{ye_parser_model_2025}. The GitHub release asset is the recommended source for scripted download, whereas the Zenodo record serves as the permanent archival reference. A correctly installed model directory should contain \path{config.json}, \path{tokenizer_config.json}, and a weight file (\path{model.safetensors} or \path{pytorch_model.bin}). Installation can be verified with \texttt{make model-check} (Section~\ref{sec:model_setup}). A Hugging Face mirror of the parser model is also available for download convenience\footnote{\url{https://huggingface.co/boshuai1/c2q-parser-codebert}}.

\section{Execution Routes and Setup}
\label{sec:paths_setup}

\subsection{Recommended Routes}
\label{sec:recommended_paths}

\textbf{Primary reproduction route.} The source-checkout installation in Section~\ref{sec:primary_path}, combined with the parser model setup in Section~\ref{sec:model_setup}, is the main route for reproducing the paper-backed executable workflows.

\textbf{Lowest-overhead sanity check.} The Docker route in Section~\ref{sec:docker_path} provides a model-free JSON smoke check without requiring a local Python environment. The Docker image minimises host-environment variability, supports clean-machine validation, and provides the most controlled initial functional route, consistent with the ACM recommendation for reproducible artifact packaging.

\textbf{PyPI functional route.} The PyPI route in Section~\ref{sec:pypi_install} installs the \texttt{c2q-framework} package for lightweight CLI and API use on Python~3.12--3.13 without cloning the repository. It provides the \texttt{c2q-json} command-line entry point for small JSON examples. It does not support the \texttt{make}-based reproduction workflow, and the parser model must still be downloaded and installed separately for any model-backed use.

\textbf{Suggested first-pass workflow.}
For a lowest-barrier initial check, evaluators may start with the Docker route in Section~\ref{sec:docker_path}, for example via \texttt{make docker-reproduce-json-smoke}. This provides a model-free sanity check in an isolated environment and is intended for preliminary functional validation only.

For actual reproduction of the paper experiments, the primary route is the GitHub source-checkout path in Section~\ref{sec:primary_path}. A practical progression is to first validate the model-free JSON pipeline, then install and verify the parser model, then run the model-backed smoke path, and finally execute the experiment-backed targets as needed. The corresponding commands are documented in Sections~\ref{sec:primary_path}, \ref{sec:model_setup}, and \ref{sec:reproduction_commands}.

\subsection{Docker Preliminary Route}
\label{sec:docker_path}
This route assumes that Docker and Docker Buildx are available; the latter can be checked with \texttt{docker buildx version}. On Linux or WSL, Docker commands may require \texttt{sudo} unless the user account has Docker daemon access. For the lowest-overhead model-free check:

\begin{cmdbox}
git clone https://github.com/C2-Q/C2Q.git
cd C2Q
make docker-build
make docker-reproduce-json-smoke
\end{cmdbox}

This route does not require the parser model and eliminates host-environment variability, making it the recommended starting point for clean-machine functional validation. If Docker requires elevated privileges, the same commands can be run as \texttt{sudo make docker-build} and \texttt{sudo make docker-reproduce-json-smoke}. After model installation, model-dependent Docker checks such as \texttt{make docker-smoke} are also available.

\subsection{Source-Checkout Installation}
\label{sec:primary_path}
This route requires Python~3.12 or Python~3.13. Confirm that \texttt{python3.12 --version} or \texttt{python3.13 --version} succeeds before creating the virtual environment.
\begin{cmdbox}
git clone https://github.com/C2-Q/C2Q.git
cd C2Q
python3.12 -m venv .venv   # or: python3.13 -m venv .venv
source .venv/bin/activate
python -m pip install --upgrade pip
python -m pip install -e ".[dev]"
\end{cmdbox}

For the model-free JSON route, evaluators may proceed directly to the JSON smoke commands in Section~\ref{sec:reproduction_commands}. For the model-backed Python-code route, the parser model should be installed first (Section~\ref{sec:model_setup}), after which the environment and model availability can be checked with:

\begin{cmdbox}
make doctor
\end{cmdbox}

A passing \texttt{make doctor} run reports that Python, \texttt{pdflatex}, and the parser model are all available for the model-backed route. If \texttt{pdflatex} is not found, report generation will fail; install a \TeX{} distribution (e.g., TeX~Live or MiKTeX) before proceeding.

\subsection{Windows PowerShell Route}
\label{sec:windows_path}
This secondary setup variant is provided for Windows users who prefer PowerShell; the primary reproduction path in this report remains the Unix-like source-checkout route as introduced in Section~\ref{sec:primary_path}. 

\begin{cmdbox}
py -3.12 --version
git clone https://github.com/C2-Q/C2Q.git
cd C2Q
py -3.12 -m venv .venv
.\.venv\Scripts\Activate.ps1
python -m pip install --upgrade pip
python -m pip install -e ".[dev]"
\end{cmdbox}

The same commands can be run with \texttt{py -3.13} when Python~3.13 is the chosen interpreter. If script activation is blocked by execution policy:

\begin{cmdbox}
Set-ExecutionPolicy -Scope Process Bypass
.\.venv\Scripts\Activate.ps1
\end{cmdbox}

\subsection{PyPI Installation}
\label{sec:pypi_install}

The PyPI package is intended for lightweight CLI and API use without cloning the repository, and is supported on Python~3.12--3.13 for this release. It does not support the \texttt{make}-based reproduction workflow.

\begin{cmdbox}
python -m pip install --upgrade pip
python -m pip install --upgrade c2q-framework
c2q-json -h
\end{cmdbox}

A minimal model-free JSON input has the following form:

\begin{cmdbox}
{
  "family": "MIS",
  "goal": "find a maximum independent set of the graph",
  "description": "Minimal MIS example",
  "instance": {
    "graph_rep": "edge_list",
    "graphs": {
      "G1": [[0, 1], [1, 2], [2, 3]]
    }
  }
}
\end{cmdbox}

With this input stored as \path{mis_01.json}, run:

\begin{cmdbox}
c2q-json --input mis_01.json
\end{cmdbox}

The expected result is a PDF report generated in the current working directory; no parser model is required. For any model-backed use via PyPI, the parser model archive must still be downloaded separately and extracted manually; it is not installed automatically by any PyPI extra.

\subsection{Parser Model Setup}
\label{sec:model_setup}

The parser model is required for Python code-input runs. Install it with:

\begin{cmdbox}
make model-setup
\end{cmdbox}

This command unpacks the model into the default path \path{src/parser/saved_models_2025_12/}. By default it looks for a local archive in common locations such as \path{~/Downloads/saved_models_2025_12.zip}; if none is found, it attempts to download from the configured GitHub release URL. If the archive has been downloaded manually:

\begin{cmdbox}
make model-setup MODEL_ARCHIVE=/path/to/saved_models_2025_12.zip
\end{cmdbox}

An optional convenience command for automated download:

\begin{cmdbox}
make model-download
\end{cmdbox}

Verify the installation with:

\begin{cmdbox}
make model-check
\end{cmdbox}

A manual installation helper is also provided:

\begin{cmdbox}
python tools/setup_model.py \
  --archive /path/to/saved_models_2025_12.zip \
  --model-path src/parser/saved_models_2025_12
\end{cmdbox}

If the model is not at the default path, the variable name depends on how you invoke the framework: \texttt{make}-based targets use \texttt{MODEL\_PATH}, while direct test invocations use \texttt{C2Q\_MODEL\_PATH}. In both cases, set it to the unpacked model directory (e.g., \path{src/parser/saved_models_2025_12}).

\section{Generating the Computational Evidence}
\label{sec:computational_evidence}
This section describes the repository-based source-checkout workflow for generating the computational evidence. Unless stated otherwise, the documented \texttt{make}-based commands in this section are intended to be run from a cloned GitHub checkout of the artifact, following the setup in Section~\ref{sec:primary_path}. Most computational results reported in the paper can be regenerated locally through this source-checkout route using the documented \texttt{make} targets, while Experiment~1 is supported primarily through released training materials and a pretrained model, and real-hardware validation remains access-dependent. For readers who prefer not to re-run the full workloads, the corresponding archived outputs are available in the Zenodo data record~\cite{ye_khan_c2q_dataset_2025} for direct inspection. A Hugging Face mirror of the dataset is also available for convenience.\footnote{\url{https://huggingface.co/datasets/boshuai1/c2q-dataset}} The Zenodo data record remains the primary archival source referenced in this report.

\paragraph{Simulator and hardware boundary.}
The executable RCR commands use Qiskit Aer as the controlled reproduction backend~\cite{qiskit_aer_docs_2025}. The real-hardware runs reported in the paper are archived as deployment evidence and can be re-executed only when comparable IBM Quantum and Helmi access is available~\cite{ibm_quantum_docs_2026,csc_helmi_specs_2026}. Aer and real-hardware outputs should therefore be compared at the workflow, circuit, report, and decoded-result level, not as identical measurement distributions.

\paragraph{Simulator and hardware boundary.}
The executable RCR commands use Qiskit Aer as the controlled reproduction backend. The real-hardware runs reported in the paper are archived as deployment evidence because access, queue state, calibration, topology, and noise vary over time on IBM Brisbane and Helmi. Accordingly, Aer and real-hardware outputs should be compared at the workflow, circuit, report, and decoded-result level, not as identical bitstring distributions or measurement histograms.

\subsection{Basic Validation}
\label{sec:basic_validation}

Before running the longer targets, we recommend starting with the default test suite:

\begin{cmdbox}
PYTHONPATH=. pytest
\end{cmdbox}

This tier does not require the parser model. Once the model is installed, a quick end-to-end pipeline check can be run with:

\begin{cmdbox}
make smoke
\end{cmdbox}

A successful \texttt{make smoke} run writes \path{artifacts/smoke/summary.json}. The smoke summary should describe a MaxCut instance with a 4$\times$4 QUBO matrix and a 4-qubit QAOA circuit, confirming that the local model-backed pipeline is functioning correctly.

To run the full model-dependent test tier:

\begin{cmdbox}
C2Q_MODEL_PATH=src/parser/saved_models_2025_12 \
PYTHONPATH=. pytest -m model -p no:warnings
\end{cmdbox}

Equivalently:

\begin{cmdbox}
make verify-model
\end{cmdbox}

\subsection{Paper Reproduction Commands and Archived Materials}
\label{sec:reproduction_commands}

\textbf{Model-free routes:}
\begin{cmdbox}
make reproduce-json-smoke
make reproduce-json-paper
make recommender-maxcut
\end{cmdbox}

\textbf{Model-required routes:}
\begin{cmdbox}
MODEL_PATH=src/parser/saved_models_2025_12 make reproduce-smoke

MODEL_PATH=src/parser/saved_models_2025_12 make reproduce-paper

MODEL_PATH=src/parser/saved_models_2025_12 make validate-dataset
\end{cmdbox}

The JSON routes exercise the model-free, schema-based input path. The Python routes and dataset validation require the parser model. All outputs from these targets are written under \path{artifacts/}~\cite{C2Q_github_2025}.

\textbf{Runtime notice.} \texttt{make reproduce-paper} produces up to 434 reports and runs for approximately ten hours on the reference test machine (Apple M1 Max). \texttt{make reproduce-json-paper} covers the full JSON example set and takes around two hours. Neither is suitable as a quick check. For initial verification, use the smoke routes instead. The JSON smoke subset covers one example each for \texttt{ADD}, \texttt{Factor}, \texttt{MaxCut}, and \texttt{MIS}.

The benchmark inputs are in two places: Python programs in \path{src/parser/python_programs.csv} (with \path{src/parser/data.csv} as backup), and JSON specifications under \path{src/c2q-dataset/inputs/json_dsl/} and \path{src/c2q-dataset/inputs/json_dsl_smoke/}. The corresponding archived reports and outputs are preserved in the Zenodo data record~\cite{ye_khan_c2q_dataset_2025} for comparison.

Table~\ref{tab:reproduction-mapping} maps the paper's three experiments to the corresponding executable and archival materials, including the reproduction mode for each experiment. Experiment~1 covers encoder evaluation (RQ1), Experiment~2 covers deployment and hardware recommendation (RQ2), and Experiment~3 covers full end-to-end workflow validation (RQ3). Experiment~1 is represented through the released training notebook, intermediate results, and pretrained model; it is not packaged as a single-command target because full parser retraining is outside the lightweight reproduction path.

\FloatBarrier
\begin{table*}[!ht]
\scriptsize
\centering
\caption{Mapping from the paper's experiments to released executable and archival materials. Experiment numbers follow those used in the paper~\cite{ye2025c2}. \textit{Exec.\ route} refers to the \texttt{make} target or notebook used to generate evidence; \textit{Archival material} refers to the corresponding preserved inputs and outputs in the Zenodo data record~\cite{ye_khan_c2q_dataset_2025}. Model~dep.\ indicates whether the parser model is required. \textit{Repro.\ mode} characterises the degree to which results can be independently regenerated.}
\label{tab:reproduction-mapping}
\begin{tabularx}{\textwidth}{p{0.8cm} p{2.2cm} p{1.5cm} p{2.6cm} p{1.8cm} X}
\toprule
\textbf{Exp.} & \textbf{Paper purpose (RQ)} & \textbf{Model dep.} &
  \textbf{Executable route} & \textbf{Repro.\ mode} &
  \textbf{Archival material (Zenodo) and key repository files} \\
\midrule
Exp.~1 &
  Encoder eval.\ (RQ1): parser classification, QCF translation, circuit gen. &
  No (notebook-driven; model for inspection only) &
  Notebook \texttt{parser\_train\_results} \texttt{\_12\_1.ipynb}
  in \texttt{src/parser/};
  model at \texttt{src/parser/saved\_models\_2025\_12/} &
  Inspection \& partial validation &
  Zenodo model record~\cite{ye_parser_model_2025}: model archive;
  \texttt{src/parser/python\_programs.csv},
  \texttt{src/parser/data.csv},
  checkpoints under \texttt{src/parser/results/} \\
\midrule
Exp.~2 &
  Deployment eval.\ (RQ2): hardware recommender across devices &
  No &
  \texttt{make recommender-maxcut} &
  Fully executable &
  Zenodo data record~\cite{ye_khan_c2q_dataset_2025}: recommender outputs;
  \texttt{artifacts/recommender\_maxcut/raw\_csv/}
  (\texttt{errors\_wide.csv}, \texttt{times\_wide.csv},
  \texttt{prices\_wide.csv}, plot PDFs);
  \texttt{artifacts/recommender\_maxcut/algorithm1/}
  (\texttt{winners.csv}, \texttt{details.csv}) \\
\midrule
Exp.~3 &
  Full workflow validation (RQ3): Python and JSON paths &
  Yes (Python); No (JSON) &
  \texttt{make reproduce-smoke} /
  \texttt{make reproduce-json-smoke};
  full-scale: \texttt{make reproduce-paper} /
  \texttt{make reproduce-json-paper} &
  Fully executable on simulator; real-hardware outputs archived and access-dependent &
  Zenodo data record~\cite{ye_khan_c2q_dataset_2025}: archived reports;
  inputs under \texttt{src/c2q-dataset/inputs/json\_dsl/} and
  \texttt{src/c2q-dataset/inputs/json\_dsl\_smoke/};
  usability materials under \texttt{src/c2q-dataset/usability\_comparison/};
  outputs under \texttt{artifacts/reproduce/} \\
\midrule
Support &
  Dataset \& parser validation (supports Exp.~1, 3) &
  Yes &
  \texttt{make validate-dataset} &
  Fully executable &
  Zenodo data record~\cite{ye_khan_c2q_dataset_2025}: validation outputs;
  \texttt{artifacts/parser\_validation/implementation/}
  (\texttt{snippet\_metrics.csv}, \texttt{family\_summary.csv},
  \texttt{syntax\_failures.csv});
  \texttt{artifacts/parser\_validation/diversity/}
  (\texttt{algorithm\_diversity\_summary.csv},
  \texttt{summary\_by\_tag.csv},
  \texttt{algorithm\_signals\_per\_instance.csv}) \\
\bottomrule
\end{tabularx}
\end{table*}

\subsection{Minimum End-to-End Checks}
\label{sec:min_examples}

\textbf{Repository-based smoke route.} The quickest check is the JSON smoke route:

\begin{cmdbox}
make reproduce-json-smoke
\end{cmdbox}

This route is model-free and writes smoke-scale JSON reports to \path{artifacts/reproduce/json/smoke/}. Running this check before any model-dependent route confirms that the JSON pipeline and \texttt{pdflatex} installation are working correctly.

\textbf{PyPI-based functional check.} Section~\ref{sec:pypi_install} gives the corresponding command-line check via the PyPI package using a small JSON input.

\section{Inspecting and Analyzing Results}
\label{sec:results_inspection}

When running the artifact for the first time, we suggest checking three things:

\begin{enumerate}[leftmargin=*]
  \item the chosen installation steps complete without errors, and for the model-backed route, \texttt{make doctor} passes after parser-model installation;
  \item the selected smoke command completes without missing-model or missing-tool errors; and
  \item the expected output directory is populated with the generated reports or summaries.
\end{enumerate}

Table~\ref{tab:expected_outputs} lists each command with its model dependency, output location, and a concrete indicator of a passing run. Locally generated outputs can be compared against the archived materials in the Zenodo data record~\cite{ye_khan_c2q_dataset_2025}.

\FloatBarrier
\begin{table*}[!ht]
\centering
\caption{Documented execution routes, model dependence, output locations, and expected passing indicators. Commands marked $\dagger$ are time-consuming and are not suitable as initial checks; see the runtime notice in Section~\ref{sec:reproduction_commands}.}
\label{tab:expected_outputs}
\begin{tabularx}{\textwidth}{p{3.3cm} p{1.6cm} p{3.0cm} p{2.4cm} X}
\toprule
\textbf{Command} & \textbf{Model req.} & \textbf{Output location} & \textbf{Experiment}
  & \textbf{Passing indicator} \\
\midrule
\texttt{make smoke} & Yes & \path{artifacts/smoke/summary.json} & Local check &
  \texttt{summary.json} describes a MaxCut QUBO matrix and 4-qubit QAOA circuit \\
\texttt{make reproduce-json-smoke} & No & \path{artifacts/reproduce/json/smoke/} &
  Exp.~3 (RQ3) &
  Four PDF/JSON reports generated (ADD, Factor, MaxCut, MIS) \\
\texttt{make reproduce-smoke} & Yes & \path{artifacts/reproduce/smoke/} & Exp.~3 (RQ3) &
  Smoke-set reports generated without error \\
\texttt{make reproduce-json-paper}$^\dagger$ & No &
  \path{artifacts/reproduce/json/paper/} &
  Exp.~3 (RQ3) & Full JSON benchmark reports; $\approx$2~h \\
\texttt{make reproduce-paper}$^\dagger$ & Yes & \path{artifacts/reproduce/paper/} &
  Exp.~3 (RQ3) & 434 benchmark reports; $\approx$10~h \\
\texttt{make recommender-maxcut} & No & \path{artifacts/recommender_maxcut/} &
  Exp.~2 (RQ2) &
  \texttt{winners.csv} selects Quantinuum H1/H2 under default weights \\
\texttt{make validate-dataset} & Yes & \path{artifacts/parser_validation/} &
  Supp.\ Exp.~1, 3 &
  \texttt{snippet\_metrics.csv} and \texttt{family\_summary.csv} populated;
  diversity summaries generated \\
\texttt{make verify-model} & Yes & local test output & Model validation &
  All model-tier pytest tests pass \\
\bottomrule
\end{tabularx}
\end{table*}

\section{Conclusion}
\label{sec:conclusion}

The framework source code is on GitHub~\cite{C2Q_github_2025} and serves as the primary executable artifact; evaluation data and archived outputs are preserved on Zenodo~\cite{ye_khan_c2q_dataset_2025}, and the parser model is archived separately~\cite{ye_parser_model_2025}. For most reproduction purposes, a source checkout under Python~3.12 or Python~3.13 with the parser model installed is the right starting point; Docker is the quickest way to confirm the JSON pipeline works before committing to the full setup, and PyPI covers lightweight CLI use.

Experiment~1 relies on the released training notebook and pretrained model rather than a single \texttt{make} command, reflecting the cost of full parser retraining. Experiments~2 and the simulator-based routes of Experiment~3 are fully executable from the released source, with concrete passing indicators documented in Table~\ref{tab:expected_outputs}. Real-hardware validation on IBM Brisbane and Finland's Helmi requires device access and cannot be independently replicated in the same sense as the simulator routes; the Qiskit Aer paths reproduce the controlled workflow, while archived real-hardware outputs document representative deployment behavior under time-specific device conditions. Known limitations on run time and platform coverage are documented in \path{STATUS.md} in the repository~\cite{C2Q_github_2025}. These include the runtime of full-scale benchmark reproduction, the separate installation of the parser model, and the access-dependent nature of real-hardware validation.

\bibliography{ref}{}
\bibliographystyle{ACM-Reference-Format}
\end{document}